\documentclass{emulateapj}

\newcommand\Exp[1]{\ensuremath{\times10^{#1}}}

\newcommand\um{\ensuremath{\mathrm{\ \mu{m}}}}

\shorttitle{IRCAL Polarimetry}

\begin{document}

\title{The IRCAL Polarimeter: \\Design, Calibration, and Data Reduction\\
for an Adaptive Optics Imaging Polarimeter}

\author{Marshall D. Perrin\altaffilmark{1,2}, James R.
Graham\altaffilmark{2}}
\email{mperrin@astro.berkeley.edu}
\affil{Astronomy Department, University of California, Berkeley, CA 94720}

\and

\author{James P. Lloyd}
\affil{Astronomy Department, Cornell University}

\altaffiltext{1}{Present address: 430 Portola Plaza, UCLA, Los Angeles
CA 90095}
\altaffiltext{2}{NSF Center for Adaptive Optics. UC Santa Cruz, 1156
High Street, Santa Cruz CA 95064}

\begin{abstract}

We have upgraded IRCAL, the near-infrared science camera of the Lick
Observatory adaptive optics system, to add a dual-channel imaging
polarimetry mode.  This mode uses an optically contacted YLF
(LiYF$_4$) Wollaston prism to provide simultaneous images in
perpendicular linear polarizations, providing high
resolution, high dynamic range polarimetry in the near infrared.  We
describe the design and construction of the polarimeter, discuss
in detail the data reduction algorithms adopted, and 
evaluate the instrument's on-the-sky performance. The IRCAL polarimeter is capable of
reducing the stellar PSF halo by about two orders of magnitude,
thereby increasing contrast for studies of faint 
circumstellar dust-scattered light. We discuss the various factors
that limit the achieved contrast, and present lessons applicable
to future high contrast imaging polarimeters.

\end{abstract}


\section{Introduction}\label{intro}

             In recent years, there has been a surge of interest in
   combining polarimetry with high angular resolution
   observing techniques, such as adaptive optics (AO). 
   This convergence is motivated both by polarimetry's 
   diagnostic potential for studies of circumstellar dust, and by its
ability to overcome atmospheric speckle noise. 
   
    For spatially resolved circumstellar disks, the polarization of dust-scattered light can 
   be used to constrain the nature of the scattering bodies, providing
   insights into dust grain structure and processing
   \citep[e.g.][]{2000ApJ...536L..89S,2004MNRAS.352.1347L,2007ApJ...654..595G}.
   The Hubble Space Telescope can provide high angular
   resolution polarimetry with very high contrast
   \citep{2006hstc.conf..153H}, but the larger
   diameters of ground-based telescopes yield superior angular
   resolution. As currently planned, the James Webb Space Telescope will not include any polarimetry capabilities.
   Thus the highest angular resolution near-infrared polarimetry in
   the foreseeable future must be obtained with AO.

   A key challenge in AO
   observations is overcoming the presence  
   of an extended, temporally-variable speckle halo in the point spread function (PSF), that limits sensitivity
   for faint material near bright stars \citep{Racine99,2004ApJ...612L..85A,2006ApJ...637..541F}.
   Because atmospheric turbulence does not polarize speckles, while
   dust-scattered light is strongly polarized,
   differential polarimetry
   can separate them to reach
   the fundamental photon noise limit for detection of faint material \citep[e.g.][]{2001ApJ...553L.189K}.

   Seeking to obtain high-contrast polarimetry of 
   dust around young stars, we developed an imaging polarimetry mode
   for IRCAL \citep{Lloyd00}, the InfraRed CAmera of 
   the Lick 3-m Shane telescope and its laser guide star AO system 
   \citep{max97,2002SPIE.4494...19B}.
    IRCAL's differential polarimetry mode 
    uses a yittrium lithium fluoride (YLF) Wollaston prism analyzer and rotating half-wave
    plate modulator to reduce speckle halos by two orders of magnitude 
    compared to direct AO imaging with the same instrument. 
    We have already published several studies of Herbig
    Ae/Be stars using this instrument, which included brief descriptions of
    the instrument itself
    \citep{Perrin2004Sci,Perrin2004SPIE,Perrin2006PDS144}. This
    present paper
    describes in greater detail the design and performance of the
    IRCAL polarimeter, and the data reduction methods adopted. Our goals are
    both to provide a detailed instrumental reference for current
    observations, and to present lessons learned that may be relevant
    for future AO polarimeters.

The structure of this paper is as follows: After briefly
restating some polarimetry theory (\S \ref{basics}), we
describe the design and construction of the IRCAL polarimeter (\S
\ref{design}) and present the results of various engineering tests of
its performance (\S \ref{tests}). We then discuss our on-the-sky
observation and calibration methods (\S
\ref{observing})
followed by the
data reduction pipeline (\S \ref{pipeline}). In 
\S \ref{performance} we evaluate the instrument's achieved contrast.
We close with a few examples of astronomical data
taken with the polarimeter, followed by a brief discussion
and conclusion (\S \ref{conclusions}).

\section{Theory for High Contrast Polarimetry} 
\label{basics}

\subsection{A Review of Polarimetry Fundamentals}

\citet{Tinbergen1996} and \citet{2002apsp.conf..303K}
provide excellent introductions to astronomical polarimetry, while 
\citet{2005ASPC..343.....A} summarizes the recent
state of the art. 

We briefly repeat here a few basic results to establish notation.
The polarization of light is usually represented by the Stokes
vector $[I, Q, U, V]$ \citep{stokes1852,chandrasekhar1946}.
The usual astronomical convention is for the $+Q$
direction to be oriented north-south, and $+U$ northeast-southwest, with angles increasing
counterclockwise from north to east.  
Linear polarization can also be expressed in terms of polarized
intensity,
$P = \sqrt{Q^2+U^2}$, and position angle $ \theta =  \frac{1}{2}\arctan({U}/{Q})$.
For astrophysical
 situations involving the scattering of light,
 circular polarization is
 usually (though not always) small compared to linear polarization,  so
 we shall generally drop Stokes $V$.
The normalized polarized intensity $P/I$ is referred to as the degree of polarization, polarization fraction or percent
polarization. Notation is not always consistent: some
authors use $P$ to refer to polarized intensity while others use it
for degree of polarization. In this work, capital $I, Q, U, V, P$ will always
refer to intensities (e.g. with units of janskies or Jy arcsec$^{-2}$), not normalized quantities. 

Visible and infrared astronomical detectors are relatively
insensitive to polarization, so to measure polarization
we must encode it
in variations of total intensity. The simplest method uses a rotatable 
linear
polarizer to allow only one polarization to reach the
detector.  To fully recover the three unknowns (Stokes $I,Q,U$)
requires measurements from at least three suitably-chosen angles.
This method has the disadvantages that (a) half the light is thrown
away, and (b) Stokes parameters are obtained from subtraction of non-simultaneous images, so atmospheric variations
cause spurious apparent polarization. This is
particularly a problem for AO observations, which have complex and time-variable PSFs.  

Dual channel polarimetry avoids this by obtaining simultaneous
measurements of perpendicular polarizations.  A Wollaston prism (\S \ref{wollsection}) placed in the
collimated beam of a conventional focal-reducer style camera produces 
two perpendicularly-polarized images of a source.  The resulting difference image should be
immune to variations in seeing or atmospheric transparency. 
A polarization modulator (such as a rotating half-wave plate)
is required to measure both Stokes $Q$ and $U$. Further modulations that swap polarization
states between the two sides of the detector can be used to minimize the
effects of flat fielding and other instrumental errors
\citep{2001ApJ...553L.189K,2006PASP..118..146P}. 
Carefully designed dual-beam, non-AO polarimeters can be made
remarkably robust against both instrumental and atmospheric
systematics, resulting in sensitivities to polarization fractions as low as $10^{-6}$ 
\citep[e.g.,][]{1987Natur.326..270K, 2006PASP..118.1305H}.

\subsection{High Contrast Differential Imaging Polarimetry}

Before we proceed into the details of the IRCAL polarimeter, we must
first establish the instrument's scientific aims.  Fundamentally,
\textit{our goal is to measure spatially varying polarization from
resolved circumstellar dust. No attempt is made to measure the total
net polarization of any unresolved source}. 
An AO polarimeter is far from an optimal tool for measuring net
polarization, due
to the complicated optical train and its inherent instrumental
polarization. (While modulation can eliminate the instrumental
polarization effects of optics after the modulator, for AO systems it
is typically impractical to locate the modulator before the entire AO
optical system.) 
For point sources, much more accurate and precise measurements are
possible using ``classical'' dual-beam, non-AO polarimeters incorporating on-axis optics
with low instrumental polarization or modulators located very early in
the optical train
\citep[e.g.][]{2005AAS...20717316P,2006PASP..118.1305H,2007PASP..119.1126M}.

While such instruments can be made extremely robust against
atmospheric and flat field effects, the situation is somewhat more complicated for AO imaging polarimeters
that attempt to resolve polarized structure hidden within the PSF
halo.   Detector limitations inevitably
constrain us to modulate more slowly than the atmospheric timescale.
Any residual image motion not corrected by the AO system must be 
compensated for by registering images prior to subtraction.
Because of this, imaging polarimeters remain sensitive both to
uncertainties in the flat field and errors in the registration
process. As we discuss below, these factors limit the achieved contrast of our instrument.

Furthermore, while AO differential polarimetry does produce a
polarized intensity image $P$ that is robust against atmospheric
errors, the total intensity image $I$ is identical to that from
regular AO imaging, including the PSF speckle halo. 
Thus $P/I$ will be a strongly biased quantity, giving in practice
only a lower limit to the true polarization fraction.  This situation
will improve with the next generation of AO polarimeters (e.g. those
in the SPHERE and GPI instruments; \citealp{2006Msngr.125...29B,2006SPIE.6272E..18M}) which will benefit from greatly improved AO PSF
quality and coronagraphic suppression of the PSF halo.

\section{Design \& Implementation } \label{design}
\begin{figure*}
\begin{center}
     \includegraphics[width=4in]{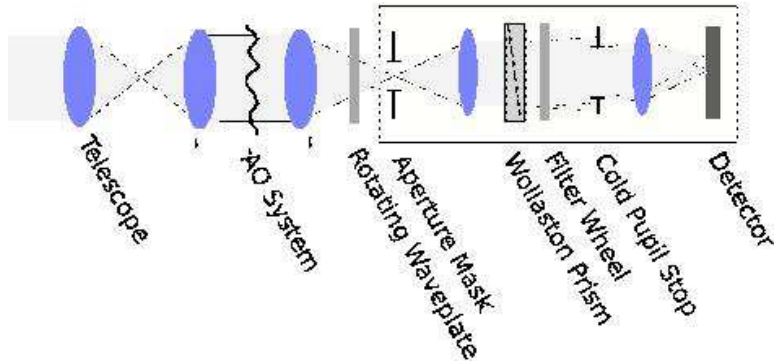}
\end{center}
    \caption{\label{schematic}  A simplified schematic of the IRCAL polarimeter, 
    not to scale but indicating the order of the various optical
    components. The box represents IRCAL's vacuum dewar. Labeled in
    bold are the polarimetry components added to the original AO system and camera: the
    rotating waveplate, the aperture mask, and the Wollaston prism mounted in
    the first of IRCAL's two internal filter wheels. The 
    Wollaston prism splits the collimated beam, forming on the
    detector
    two simultaneous images with perpendicular polarizations. Rotation
    of the waveplate modulates the intensity of the two images, allowing the measurement of Stokes $Q$ and
    $U$. }
\end{figure*}

\subsection{General Design Considerations}

\label{wollsection}

A Wollaston prism \citep{Wollaston1820}
deflects perpendicular
polarizations oppositely, with the
angle of each output beam approximately equal to
    \[
    \delta \simeq \Delta{n} \tan \alpha     \]
    where $\alpha$ is the prism angle and $\Delta{n}$ is the birefringence.
    If such a prism is placed conjugate to the telescope pupil, it will split the 
    incident beam to create two 
    polarized images on the final detector plane. The Lagrange invariant implies that 
    each of these images 
    will have an
    apparent sky-projected deflection of
    \begin{equation} \label{deflectioneq}
	\delta_\mathrm{projected} \simeq  \left( \frac{D_\mathrm{pupil}}{D_\mathrm{tel}} \right)
 \Delta{n} \tan
	\alpha     \end{equation}
    where $D_\mathrm{pupil}$ and $D_\mathrm{tel}$ are the pupil image
     and telescope diameters (5 mm and 3 m respectively in our
     case).  
     The polarimetric field of view is maximized if the total separation
     $2\delta_\mathrm{projected}$ is half the detector's
     total field of view. For
     IRCAL, this field of view is 19.4\arcsec, implying an optimal
     $\delta_\mathrm{projected} \sim 4.9''$.  Thus once we have
     chosen our material (and therefore $\Delta{n}$), equation
     \ref{deflectioneq} may be solved for the necessary prism angle
     $\alpha$.

\citet{1997A&AS..123..589O} proposed using a wedged double Wollaston for
simultaneous measurement of all four polarization states, and thus simultaneous
extraction of Stokes $Q$ and $U$. However, this approach requires dividing the
pupil in half, resulting in a loss of half the angular resolving power of the system,
so the wedged double Wollaston is not an optimal solution for high resolution
imaging. The pupil division also means that the $Q$ and $U$ beams see
different portions of the atmosphere and will have PSFs at least as divergent
as if they were non-simultaneous.  Furthermore, 
obtaining a given signal-to-noise level in both $Q$ and $U$ requires just as
much total exposure time as in a regular dual-beam polarimeter.  We therefore chose to
develop a classical single-Wollaston polarimeter.

 We also chose to use a conventional rotating half-wave plate
    as the polarization modulator, instead of alternatives such as  
    liquid crystal variable retarders (LCVRs) or ferroelectric
    liquid crystals.  Ferroelectric liquid crystals are unsuitable
    given their too-fast modulation compared to typical infrared detector readout times.
             LCVRs provide spectacular levels of precision in solar polarimetry,
                            but are generally not achromatic, are difficult to obtain for
    wavelengths beyond 1.8\um, and have birefringence that is
    temperature-dependent.  In contrast, half-wave plates are
    available with excellent achromaticity and low sensitivity to
    temperature variations.  Since we desired a single 
    modulator for the entire 1-2.5\um\ band, we use an
    achromatic waveplate as our modulator.
    Figure \ref{schematic} shows the overall layout of the 
    IRCAL polarimeter including the above-mentioned components.

\subsection{Choice of material for the Wollaston prism}

\begin{figure*}
    \begin{center}
        \includegraphics[height=2.5in]{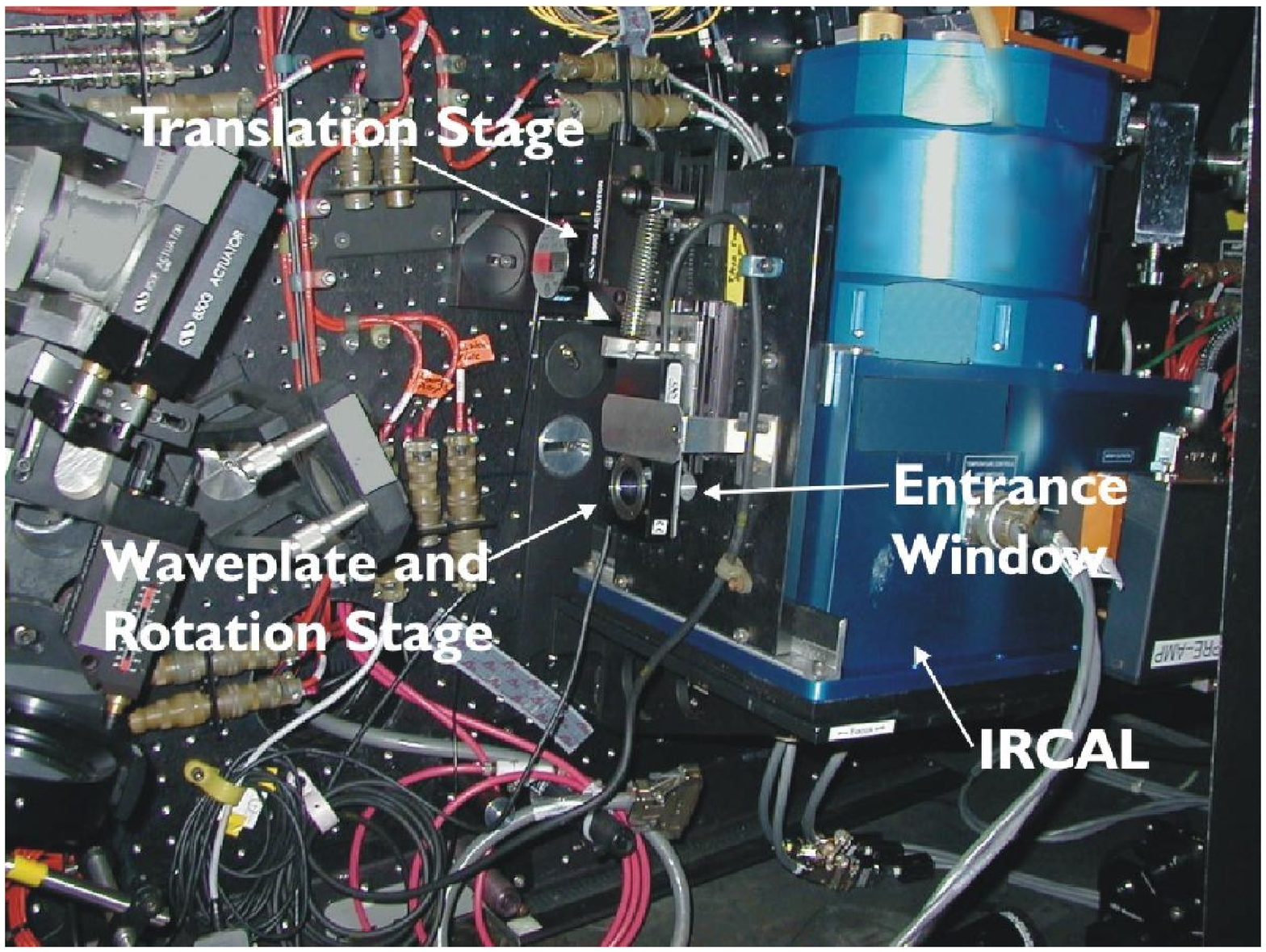}
     \includegraphics[height=2.5in]{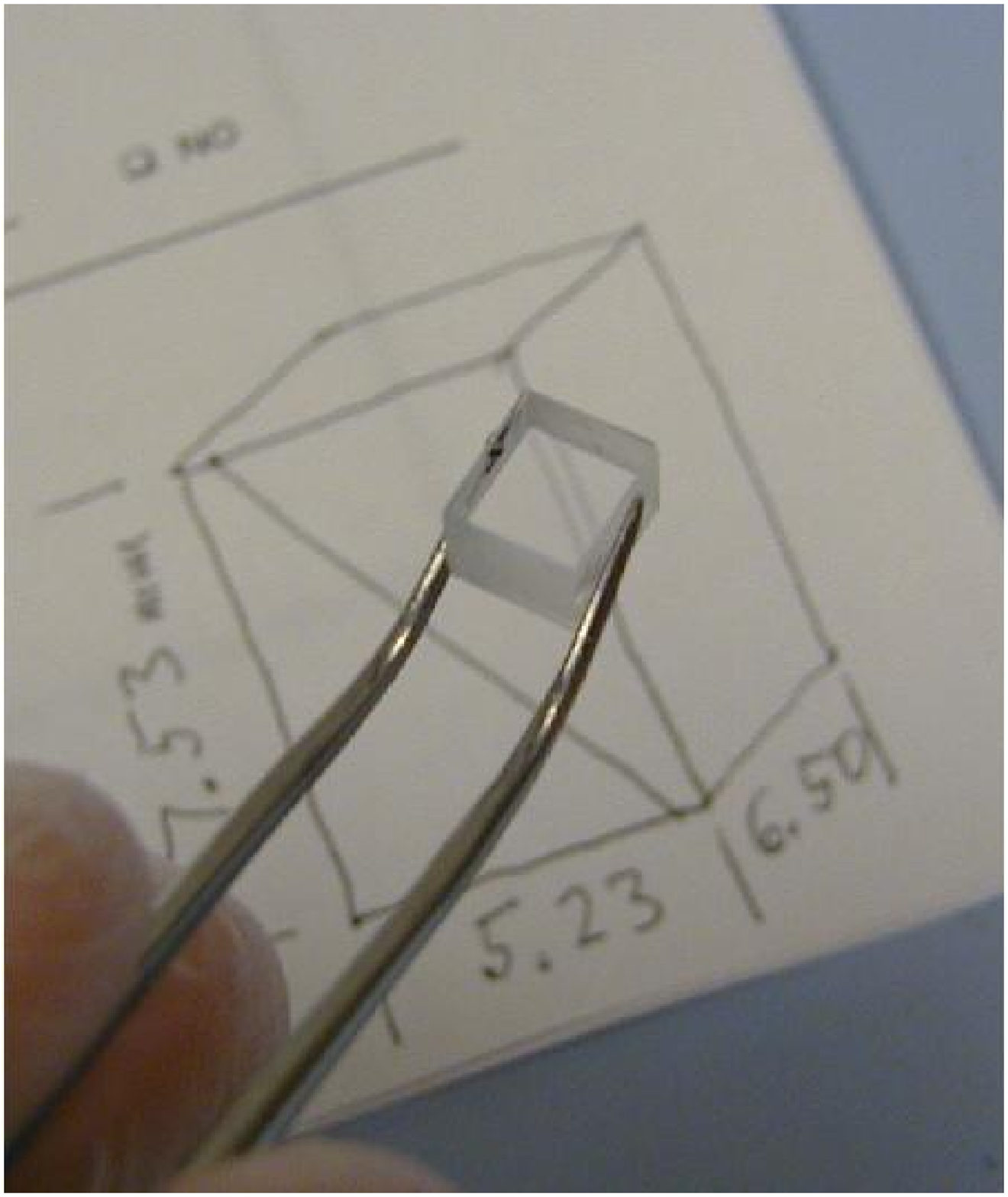}
    \end{center}
    \caption {\label{dewarphoto} 
    \textit{Left:} The IRCAL dewar mounted on the Lick AO optical bench at the Cassegrain focus of the 3-m Shane telescope, highlighting
    the waveplate mechanisms. A rotation stage allows modulation of
    the polarization using the waveplate, and a linear translation
    stage allows inserting/removing the waveplate from the beam. 
    \textit{Right:} 
    Our YLF Wollaston prism, manufactured by Onyx Optics of Dublin,
    CA.  This prism has been installed in one of IRCAL's two cryogenic filter wheels since
    March 2002.
    }
\end{figure*}

\label{materials} 

\citet{Oliva97} provides an overview of candidate birefringent materials
for infrared Wollaston prisms.
We present in Appendix A an updated list of such materials, along with recent references for their
optical properties.

For an imaging polarimeter, it is important to choose a material whose
birefringence $\Delta{n}$ does not vary much within each operating
bandpass.
A wavelength-dependent $\Delta{n}$ causes
different beam deflections for each wavelength within a bandpass,
elongating the final image.  Minimizing this effect, called 
``lateral chromatism'', 
is particularly important for a high
contrast polarimeter: not only does lateral chromatism decrease
angular resolution, it can greatly degrade PSF subtraction quality by 
oppositely distorting the two polarizations' PSFs.
Lateral chromatism may be quantified over a given
wavelength range $\lambda_1 < \lambda < \lambda_2$ by Oliva's
parameter 
       \[V_{\Delta{n}} =
       \frac{\Delta{n_{(\lambda_1+\lambda_2)/2}}}{\Delta{n_{\lambda_1}} - \Delta{n_{\lambda_2}}}
       \]
This parameter is the birefringent analogue to the Abbe number $V$,
which measures dispersion in optical materials. High values of
$V_{\Delta{n}}$ correspond to low amounts of birefringent dispersion.
The image elongation over the range $\lambda_1 -
\lambda_2$ will be $\epsilon = \delta / V_{\Delta{n}}$. By selecting a material to
maximize $V_{\Delta{n}}$, we can minimize blurring due to
lateral chromatism.

We chose the unaxial birefringent material YLF (yttrium lithium fluoride, LiYF$_4$, pronounced ``yilf''
by its users in the laser industry) because its
$\Delta{n}$ is remarkably constant with
wavelength, causing negligible 
lateral chromatism. Typically, $V_{\Delta{n}} > 100$ is considered good, corresponding to chromatic
elongation less than 0.05\arcsec\ for IRCAL, better than the diffraction
limit.  YLF exceeds this (spectacularly at $H$, with $V_{\Delta{n}} = 1290$)
and is readily available due to its widespread use in the laser industry. 
Its refractive index, 1.45, is similar to that of standard optical
components, simplifying the optical design and
manufacturing, and its birefringence, $\Delta{n}$ =
0.022, allows the necessary beam divergence to be obtained with an
reasonable thickness of YLF, roughly 4 mm. 
       
    Three similar Wollaston prisms were fabricated by Onyx Optics of
    Dublin, CA and were received at Berkeley in March 2002 (Figure
    \ref{dewarphoto}). 
    The pieces were joined
    using optical contacting and both exterior surfaces received a
    1-2.5\um\ antireflection coating.
    The 30\degr\ prism 
    angle produces a beam divergence of 1.45\degr,
    giving a sky-projected separation of 8.6\arcsec\ between the two beams 
    (close to the optimal value of half the field of view).
    All three prisms performed as expected in laboratory tests. 
    Two of the prisms exhibited minor surface defects, so we selected the third for
    astronomical use.
    That Wollaston was installed in one of IRCAL's internal filter
    wheels, in collimated space just before the cold pupil.

    A concern with any birefringent material is that the differing
    optic axes will have different thermal expansion
    coefficients.  
    For Wollaston prisms constructed by bonding two surfaces with
    perpendicular axes, temperature changes create stress
    between the two faces. 
    YLF has fairly similar expansion
    coefficients of $5\Exp{-6}\ \mathrm{K}^{-1}$ parallel and 
    $8\Exp{-6}\ \mathrm{K}^{-1}$ perpendicular to the fast axis.   
                         Our prism has tolerated
    repeated thermal cycling to liquid nitrogen temperatures with no ill
    effects apparent after five years.

One caveat about YLF is it is relatively expensive and
can be hard to obtain in large sizes, such as needed for Wollastons on
8-10 m telescopes.
The greater angular resolution of large
telescopes places a stringent constraint on lateral chromatism, 
requiring high $V_{\Delta{n}}$ materials like YLF for
diffraction-limited performance.
However, a large YLF Wollaston
was recently fabricated for the Subaru Telescope \citep{2007AAS...210.7803H}, demonstrating that
the crystal fabrication challenges can be overcome.

\subsection{Waveplate and Aperture Mask}

    We obtained an achromatic waveplate from Karl Lambrecht Corp., fabricated from MgF$_2$ and crystal quartz.
    The waveplate is optimized for 1-1.6\um, and provides $0.500\pm
    0.005$ waves of retardance over that range.  At longer
    wavelengths, the retardance differs from half a wave by about 0.05 waves averaged across the $K_s$ band. As a result the
    polarizing efficiency of the instrument is degraded slightly at
    long wavelengths, as discussed below in \S 4.
    The waveplate is installed in a Newport PR50 rotation
    stage, itself mounted on a translation
    stage for insertion into the optical path (Figure \ref{dewarphoto}).
    For practical reasons, the waveplate is located at the entrance
    window to IRCAL. Ideally, it should have been placed 
    as far upstream as possible, since modulation allows any instrumental
    polarization from post-modulator optics to be separated from
    the astronomical signal, but given
    the constraints of the AO bench that was not feasible. As a result,
    polarizing optics before the modulator, such as tilted mirrors
    and the dichroic beamsplitter, introduce a nonzero instrumental
    polarization.

    To prevent overlap of the two polarized images, 
    we installed in IRCAL's cold aperture wheel a rectangular mask
    corresponding to half the field of view,     essentially a $9\arcsec$ wide slit.     As fabricated, the mask is 
    slightly oversized, resulting in an 0.4\arcsec\ (6 pixels or 240\um) beam overlap
    in the middle of the detector. This overlap region is 
    discarded during data reduction.

\subsection{Limitations and Imperfections}

The IRCAL polarimeter's performance is limited by several factors
in its design and construction.  For instance, errors in flat fielding 
prevent completely accurate subtraction of the two polarizations.
Modulation reduces the impact of this, but
cannot eliminate the problem entirely. Even though both polarizations
are alternately measured on both sides of the detector, the images
must be shifted to register them before subtraction, which depends on
accurate flat fielding to maximize the cancellation of starlight. 
      
    A second factor that limits real-world polarimeters is
      registration and shifting errors
      \citep[e.g.][]{2004AnA...415..671A}.  IRCAL's coarse pixel
      sampling (which only Nyquist-samples the PSF at wavelengths $> 2
      \um$) exacerbates this. The undersampled nature of $J$- and
      $H$-band data limits the precision possible in registration
      and introduces aliasing that compromises image subtraction.
      Even were the sampling improved, accurately registering 
      images would still be challenging, particularly for deep
      exposures saturated on the central core.

The placement of the Wollaston prism in a filter wheel \textit{before} the cold
      pupil results in the two polarized beams
    already having diverged by the time they reach that pupil stop.
    This ``pupil shear'' causes each channel to have a slightly different effective pupil and significantly increases the non-common-path
    aberrations between the two beams.
    Traced backward through the telescope and onto the sky, each channel sees a
    slightly different column of partially-corrected atmosphere, displaced from the other by $\sim 5\%$ of the pupil diameter.
    This results in non-common-path \textit{atmospheric} wavefront
    errors, which cannot be compensated for by waveplate modulation.
    It would have been much better to place the Wollaston after
    the cold pupil, but the layout of the existing filter wheels
    didn't allow this.
 
IRCAL's aperture mask wheel is not a precision mechanism. Its positioning has low repeatability,
    and it occasionaly shifts unpredictably 
    (apparently as a result of electrical crosstalk from one
    of the other wheel mechanisms). Shifts in the aperture
    mask change the field of view and flat field for the
    observations. The filter wheel containing the Wollaston is less
    sensitive, but slight offsets from night to night change
    the relative position of the two polarized images, which the data
    reduction pipeline must deal with.

    Compared to these factors, 
    any imperfections in the polarimetry optics themselves, and/or the 
    finite precision of the waveplate rotation (0.1\degr), are 
                                 entirely negligible.  Wollaston prisms frequently achieve
      extinction ratios better than $10^{-5}$; at this time, we 
      cannot more precisely quantify the performance of
      our prism but believe that its extinction ratio and polarizing
      efficiency are excellent.

\section{Commissioning Tests} \label{tests}

    The prism was installed into IRCAL during
    2002 March, and had first light on 2002 March 29 UT. Observations then and on
    subsequent nights have allowed us to quantify the performance of the system.

    The polarimetry optics have negligible effect on
    the achieved AO image quality. 
                They do contribute additional
    non-common-path wavefront error between the science detector and 
    wavefront sensor (including about 1.3 mm of focus and
    85 nm RMS of 
    other aberrations, principally astigmatism) but by applying
    appropriate offsets to the deformable mirror, we can fully compensate
    for these aberrations. This results in internal Strehl ratios
    of $0.92\pm 0.02$, essentially identical to performance
    without the polarimetry optics\footnote{Strehl ratios are measured using an IDL-based
    PSF fitting routine developed by Bruce Macintosh and incorporated
    into the IRCAL quicklook software. There are numerous pitfalls in
    accurately measuring Strehl ratios, particularly for undersampled
    data \citep{2004SPIE.5490..504R}, so Strehl ratios quoted for
    IRCAL are uncertain by $5-10\%$ in an absolute sense}.
    The resulting orthogonal polarized PSFs appear very similar 
    but not perfectly identical, confirming the presence of
    small non-common-path wavefront errors between the two
    polarizations.

    We observed the Lick AO system's internal calibration source to
    determine the throughput of the Wollaston prism and
    waveplate, including both reflective losses and absorption.
        At longer wavelengths, YLF
    becomes more transparent, but the waveplate's antireflection
    coating performs less well. This results in a net transmission
    through both optics of $92.3\pm0.3\%$ at $J$ and $H$, decreasing to $88.5\pm0.5\%$ at
    $K_s$.

      To calibrate the polarizing efficiency, we placed in front of the AO calibration source a 1-2\um\ IR
      Polaroid from Edmund Scientific, and observed its apparent brightness as a
      function of waveplate angle, for each of $J$, $H$, and $K_s$.
      We fit a model to these data using the
      Levenberg-Marquardt least-squares method to extract the polarizing efficiency. The polarizing efficiency measured in this way is 
      $1.00\pm0.005$ for $J$ and $H$, and $0.95\pm
      0.01$ for $K_s$ band. This decreased polarizing efficiency at
      $K_s$ is due entirely to the
      waveplate's retardance differing slightly from half a wave at wavelengths
      beyond 1.8 \micron.

\subsection{Instrumental Polarization}

    Instrumental polarization of some degree is found in all
    optical systems, particularly those that depart from circular
    symmetry, such as by having oblique reflections. The Lick AO optical path involves
    many such reflections, and thus moderate instrumental
    polarization is expected. Accurately modeling instrumental
    polarization is challenging, as oxide layers on mirrors
    and other subtle imperfections can play substantial roles
    \citep{2002apsp.conf..303K}. 
      For very high contrast applications, the polarizing effects of thin metal
      films must be considered
      \citep{BornandWolf1959,2004ApJ...600.1091B}.

    We observed unpolarized
    standard stars \citep{1974psns.coll.....G,Ageorges99} to calibrate instrumental polarization.  We
    measured the apparent polarization of each standard via aperture
    photometry on the reduced data, and interpreted the apparent
    polarizations as the amount of instrumental polarization.
For all wavelengths, the derived instrumental polarizations are around 2.5\% in the
$-Q$ direction.
       Due to the Cassegrain configuration, these values are extremely stable with time,
changing $< 0.1\%$ in observations of standard stars more than a year
apart.  
Based on comparison of these measurements to a Mueller matrix model
of the optical train, the instrumental polarization appears
to be primarily caused by the 
dichroic that 
separates the visible wavefront-sensing and IR science channels. This
large contribution is unsurprising for a complex multilayer optic at oblique 
incidence.
We did
not attempt to measure circular instrumental polarization or crosstalk
between circular and linear polarizations.

The optimal method to remove instrumental polarization depends on the
science goals. 
We adopt an iterative optimizing
subtraction process that simultaneously removes linear instrumental and
interstellar polarization, described in \S \ref{red_general}.

\subsection{Pixel Scale}

\label{pixelscale}
    
Lloyd et al. (2000) previously measured IRCAL's pixel scale as
$0.0756\arcsec\pm0.0002$ pixel$^{-1}$. Subsequent observations
have allowed a refined
measurement, revealing slightly different plate scales in R.A. and
Dec. This indicates that IRCAL is anamorphic, a result not
unexpected given that its internal off-axis parabolic mirrors are not
of unit magnification. 

To quantify the anamorphism, we obtained LGS AO observations of the
core of the globular cluster M 53.  We extracted point source
coordinates using the \texttt{Starfinder} IDL package, and compared them to
coordinates derived from HST WFPC2 imaging of the same cluster
\citep{2002A&A...391..945P}. We
simultaneously fit the displacement, rotation, shear, and x- and
y-magnification between the two datasets using the MPFIT
Levenberg-Marquardt nonlinear least squares optimizer\footnote{http://cow.physics.wisc.edu/~craigm/idl/fitting.html}. This analysis
yielded plate scales of $0.0754\arcsec$ pixel$^{-1}$ in R.A.
and $0.0780\arcsec$ pixel$^{-1}$ in Dec., with statistical error $\pm
0.0005\arcsec$ pixel$^{-1}$. The detector's axes are orthogonal within our 0.5\degr\ measurement uncertainty.

To correct the anamorphism, we developed an IDL routine
\texttt{ircal\_dewarp}, based on Keck's \texttt{nirc2warp} by Antonin
Bouchez. This routine transforms a distorted input image
onto a uniform 0.040\arcsec\ grid via two-dimensional cubic spline interpolation. 

To assess the performance of this correction,
and to measure any shifts in the absolute orientation of IRCAL over time,
we routinely observe binary stars drawn
from the Suggested Calibration Targets list of the Washington Double
Stars (WDS) Orbits Catalog \citep{2001AJ....122.3472H}. The observed binary separations are 
compared to values computed from the
WDS orbital elements.
Because the WDS Orbits Catalog is itself 
imperfect (most of the binaries with suitable separations and
brightnesses have orbits graded 4
or 5, the lowest grades) some amount of discrepancy is expected. Since many entries in the WDS do
not contain estimated uncertainties, we make no attempt here to
separate out these two effects.  A conservative estimate is
that the pixel scale in post-anamorphism-correction IRCAL images is
uncertain by no more than $\sim$2.5\% in linear size, and by $\sim$1.0\degr\ in
position angle.  

The cubic spline interpolation employed is a good
approximation to the theoretically optimum sinc 
interpolation, but not perfect, particularly given 
the non-Nyquist-sampled nature of short-wavelength IRCAL data.  ``Ringing''
artifacts (the ``Gibbs phenomenon'') are sometimes seen near stellar PSF peaks in
dewarped images.
It is worth noting that the \texttt{ircal\_dewarp} routine is not strictly flux
conserving, nor does it account for the small higher-order distortions
present in IRCAL (M.~P.~Fitzgerald, private
communication). 
Attempting to characterize and compensate for these effects,
either through heuristic algorithms such as
Drizzle \citep{2002PASP..114..144F} or through more formally correct
Fourier interlacing methods (Bracewell 1978, Lauer 1999), is beyond the scope
of this paper.

\section{Observing Techniques and Astronomical Calibration}
\label{observing}

\subsection{Observing Procedure}

         Our standard observing procedure is as follows: 
    Once
    the target has been acquired, the
    observer configures settings such as filter and
    exposure time. The camera's software automatically rotates the
    waveplate and integrates at each position. 
    We use eight
    distinct waveplate positions: 0, 22.5, 45, 67.5, 90, 112.5, 135,
    and 157\degr. This provides four redundant measurements for
    each of Stokes $Q$ and $U$, reducing
     instrumental systematics
    \citep{2001ApJ...553L.189K,2006PASP..118..146P}.  To obtain the
    necessary high dynamic range, for most targets it is necessary to
    combine both short and long exposures, typically 
    between 0.5 and 90 s, depending on source magnitude. Small dithers 
    after each set of eight exposures are used to reduce
    flat field effects and/or increase field of view.

\subsection{Flat fielding } 
As flat-fielding errors are expected to be a major contributor to
systematic limits on 
the detection of faint polarized circumstellar material,
several methods were pursued for flat-fielding the polarimeter.
The IRCAL aperture mask and filter wheel positioning mechanisms have frustratingly poor
repeatability.  Because of this, the aperture mask position can only
be approximately reproduced from night to night (typically within 5-10 pixels), so it is
necessary to take flats again every night.  Some authors have advocated
computing flat field calibrations using observations of extended
sources such as planets \citep{1991PASP..103.1097K}, but such targets
are not always visible. We instead opt for the more traditional 
sky as flat calibrator.

Because the twilight sky is itself highly polarized, it is far from
ideal for flat fielding a polarimeter.  Dome flat screens are
typically only slightly better. 
To get around this problem, polarized sky or dome exposures taken
with different waveplate angles are averaged to make
``pseudo-unpolarized'' flat frames. This method only works perfectly if the sky
brightness and polarization are constant, a requirement notably not
satisfied during twilight (see \citet{Cronin:2006p303} for discussion of sky polarization during twilight and its variation with time). IRCAL's small pixel scale allows it to take
unsaturated exposures on the daylit sky hours before
sunset, at which point the sky polarization is more stable. 

From each night's set of ``pseudo-unpolarized'' frames we compute one master
median flat per wavelength, which we use for all waveplate angles. 
Errors in the final master flats are estimated to be a few parts in a
thousand. 
Some authors have
reported slight benefits from using a flat frame
computed individually for each waveplate angle \citep{Ageorges99}.
We did not investigate that approach in detail, but note
that it may offer future potential for improvement. 

The nightly flat fields are also used to measure the location on the detector of the 
two polarized subimages, using a Hough-transform 
edge detection algorithm to find the edges of the illuminated fields of
view. 
The detected masks are undersized by 5
pixels to avoid light scattered from the aperture
mask's edges.

\subsection{Sky Subtraction}

At night the
infrared sky background is primarily unpolarized, vibrationally
excited emission
from atmospheric molecules.  In dual-channel
differential polarimetry, this unpolarized background will subtract out
of the $Q$ and $U$ images. Thus there would be no need to worry about
sky subtraction, if our sole concern was the polarized images.
However, since we also want to measure total intensity for each
target, we perform the usual sky subtraction. We subtract a distinct sky
frame for each waveplate angle, obtained either from separate sky frames or as a
median of the dithered science frames with that angle.
This process also allows for the subtraction of
any polarized sky light such as scattered moonlight, but as mentioned above, this is usually
negligible.

\subsection{Polarization Angle Calibration}

    The Stokes reference frame must be oriented
    relative to
    some known reference angle. (Equivalently, we must
    determine the rotational zeropoint for the waveplate mechanism.)
    We do this using 
    the twilight sky, which is strongly linearly polarized
    perpendicular to the direction pointing toward the sun \citep{Cronin:2006p303}. 
                Using the sky as a
    calibrator allows us to use all pixels of the detector and thus
    rapidly obtain high signal to noise. In practice, the same data
    are used for flat fielding the detector and for calibrating the
    position angle\footnote{We note in passing that
    \citet{2006PASP..118..845H} report difficulties using the
    sky as a position angle calibrator in Hawaii, because of 
    specular reflection of the sun by the ocean, but we have found
    that it works adequately for our purposes at Mt. Hamilton.}.

\subsection{Photometric Calibration} 
Photometric standards
\citep[e.g.][]{PerssonStandards,2001MNRAS.325..563H} are observed
using the same procedure as science targets.
After the standard polarimetry reduction (described in \S
\ref{pipeline}), we measure their fluxes within a 7\arcsec\ diameter 
aperture, plus encircled energy curves for aperture correction.  Adaptive optics
photometry is a notoriously problematic area due to PSF
variability \citep[see][]{2006ApJ...637..541F,2006ApJ...647.1517S}. Changes in sky
conditions and guide star brightness inevitably cause wavefront correction
to vary between science target and calibration source,
causing photometric errors as the encircled energy curve changes.  We
expect an uncertainty of 5-10\% in the resulting photometry in typical
conditions.

\section{The IRCAL Polarimetry Data Reduction Pipeline}\label{pipeline}

Data reduction for imaging polarimetry has been discussed before by a
number of authors 
\citep{1991MNRAS.252P..50G,1995Ap&SS.224..395W,2001ApJ...553L.189K,PotterPhD,2006PASP..118..146P}. The reduction algorithms for
IRCAL build on these earlier works but also feature a number of
elements specific to IRCAL. 
We describe here the data reduction process 
in detail, in the hope of sharing lessons from IRCAL with other aspiring
differential polarimetrists. 
Furthermore, this is an appropriate setting to air 
dirty laundry: that
is, to highlight what \textit{doesn't} work about IRCAL polarimetry as
well as what does, and what aspects should be improved in future
designs, such as the planned polarimetry mode of the Gemini Planet
Imager.

\subsection{General Considerations}
\label{red_general}

Our fundamental goal is to use polarimetry to cancel out the
unpolarized stellar
PSF and reveal faint circumstellar material. In practice,
complications such as instrumental and interstellar polarization mean
that the star's PSF is not truly unpolarized. 
For distant or young and embedded sources, 
interstellar polarization can be substantial, at a level of a few
percent. 
While it is possible to remove interstellar polarization
based on observations of nearby stars \citep{2000AJ....119..923H}, this
is fraught with uncertainties. Not least of these is the
assumption that interstellar polarizations vary in a predictable and
uniform way between stars--
an assumption unlikely to be satisfied in the clumpy, dusty environments of
young stars.  Is there a better approach?

We can instead allow the net polarization of the star (arising from
both interstellar and instrumental polarization) to be a free
parameter that we solve for. We do this by finding the 
scaling coefficients that minimize residual starlight in 
$|Q|$ and $|U|$, following the approach used by
\citet{PotterPhD}. 
This method relies on the assumption that the total flux is
dominated by inherently unpolarized light, so that any net
polarization seen in the data is the result of instrumental or
interstellar polarization. For our science targets, this is generally
an good approximation, since scattered light is no more than a
few percent of the total. 

This approach is robust even in the presence
of scattered light. For a symmetric distribution of dust around a
star, the polarization position angle varies around the disk and the
total net polarization in a large aperture cancels to zero. 
For inclined or
nonaxisymmetric dust distributions, that cancellation begins to break
down and can introduce a systematic bias to underestimate the 
true degree of polarization.  In an extreme case, a 
100\% linearly polarized source would not be measurable with this technique at all.
But as mentioned above, we restrict our efforts to the
detection of spatially varying polarization from resolved dust. For
such systems, the uncertainty resulting from this scaling process will
be negligible compared to other factors, such as the biasing effect of
the PSF halo in Stokes $I$.

With that goal in mind---the detection of faint spatially resolved
polarization from extended circumstellar dust, and not the precise
measurement of total polarization---an IDL data reduction pipeline for
IRCAL polarimetry data has been developed. This software is available
online\footnote{\texttt{http://astro.ucla.edu/$\sim$mperrin/ircal/}}
and automatically handles the reduction, calibration, and analysis of IRCAL
polarimetry data. 
Minimal user intervention is required to produce first-pass science
reductions, though extensive options exist for overriding or
fine-tuning the pipeline's behavior.

\subsection{Initial Reduction}

Images are first dark-subtracted, then divided
by the polarization mode flat field. Sky
subtraction is performed, as described above. 
Bad pixels are identified and removed, using the combination of a 
mask indicating known permanently-bad pixels and the NIRSPEC
\texttt{fixpix} iterative bad pixel cleaning routine developed by Tom
Murphy.
At this point, the images are relatively clean, and can be
corrected for anamorphism using the \texttt{ircal\_dewarp} routine (\S
\ref{pixelscale}). 

Saturated pixels and/or pixels above the detector's linear regime are
identified by comparison with a threshold value of 280,000
electrons pixel$^{-1}$.
The saturated pixel lists for the two polarized subimages
are merged to ensure that if any pixel is marked as saturated, its
corresponding pixel in the other polarization is also marked as saturated.
Without this step, polarization artifacts would be induced later
during image subtraction.

\subsection{Registration}
Accurate registration to a fraction of a pixel 
is required to extract polarized signals very close to a star.  \citet{2004AnA...415..671A}
found that for AO polarimetry at the Very Large Telescope, random shifts of 0.2-0.5
pixel (5-14 mas) distort the polarized signal of the inner arcsecond,
and shifts of 1 pixel completely disrupt it. Long exposure images,
which probe polarization on slightly larger scales, are more robust
against misalignments. Apai et al.\ used a
``two-level Gaussian fitting procedure'' to register their data, while
\citet{PotterPhD} used an iterative shift-and-subtract
process to empirically find the shift for each image that gave the
lowest residuals. For IRCAL we use Fourier cross-correlation to subpixel
precision. Which of these
techniques is best will vary depending on instrument and
observational conditions.

The first step in image registration is to determine the offset
between the two polarized subimages on the detector, which varies
slightly from night to night due to imprecision in the filter wheel
rotation.  To measure this, the full stack of images is summed, and
the resulting left- and right-hand subimages are cross-correlated to
find their displacement.

The individual images are then registered via cross-correlation,
fit by Gaussians to subpixel precision.  
Only the \textit{left-hand} subimages are correlated, with
the same shifts (plus the beam displacement) assumed correct for the
right hand field. 
If
necessary, manual intervention is possible to tweak the automatic
registration.

An important caveat about IRCAL polarimetry is that the images are
only Nyquist sampled at $K_s$. At shorter wavelengths, high spatial
frequency PSF structure will alias to lower spatial frequencies, biasing the
correlations and derived shifts in an unpredictable and time-variable
manner. In the absence of \textit{a priori} PSF knowledge, this imposes a fundamental limit on our ability to accurately
register these images. Furthermore, the 
long exposures necessary to detect faint circumstellar dust are
frequently saturated on the central star, which limits registration
accuracy even more.  Cross-correlation still works to
some extent in this case,
using the diffraction spikes and other outer PSF structure, which is
why we chose that approach for the IRCAL pipeline.
The highest accuracy polarimetric image registration
may require the use of ``satellite
PSF'' techniques developed for coronagraphic observations
\citep{2006ApJ...647..612M,2006ApJ...647..620S}.

\subsection{Mosaicing}

All the subimages for a given polarization are then mosaiced together,
combining different exposure times via a weighted average.  
Subpixel shifting is done by Fourier interpolation. 
A separate mosaiced image $M_i$ is produced for each of four polarization angles:
$M_0, M_{45}, M_{22.5}$, and  $M_{67.5}$. These mosaics combine polarized
subimages from both sides of the detector, and images from the
redundant waveplate settings 90, 135, 112.5 and 167.5 are summed
with their counterparts (i.e. images taken with the waveplate at
90$\degr$
are mosaiced in with those taken at 0, and so on.)

The resulting mosaics are then scaled to minimize their subtracted
residuals $Q$ and $U$, according to 
\begin{eqnarray*}
    Q &=& M_0 - \alpha M_{45} + \beta \\
    U &=& M_{22.5} - \gamma M_{67.5} + \delta \\
    I &=&  \left( M_0 + \alpha M_{45} +  M_{22.5} +\gamma M_{67.5} \right)/2.
\end{eqnarray*}
The coefficients for minimizing the total power
in $|Q|$ and $|U|$ are determined by IDL's Powell nonlinear optimization
routine. 
Typical values for $\alpha$ 
and $\gamma$, the scaling factors that remove the net polarization of
the targets, are within a few percent of unity.  
The $\beta$ and $\delta$ factors correct for any residuals in the sky
background levels, and are always small. 

This order of operations---mosaicing the images, then performing the
scaled subtraction---is by no means the only way to do things. One can also
first difference the images and then mosaic the subtracted frames, or
there are also reduction algorithms that rely upon ratios rather than
differences of the two channels \citep[see][]{Tinbergen1996}. We chose
the
mosaic-then-subtract approach because it offered 
greater flexibility for dealing with missing frames from AO system
dropouts or moments of bad seeing, but the merits of the various
algorithms should be reevaluated for future instruments. 

The polarized intensity $P$ is computed from $Q$ and $U$ in the usual
manner, $P = \sqrt{Q^2+U^2}$.  
If absolute accuracy in $P$ were our goal, we would need at this
point to remove the bias introduced into $P$ by its positive definite nature, by
following one of the the standard statistical debiasing algorithms
\citep{1974ApJ...194..249W,1985A&A...142..100S,1991A&A...246..280S}.
But we skip this step for two reasons. 
First, these corrections are important primarily when
the signal to noise ratio is low ($P/\sigma_P \lesssim 2$), and
we focus our analysis only on pixels
with strong detections of polarization ($P/\sigma_P > 3$). Much more importantly, 
for AO data the bias in degree of
polarization $P/I$ from the uncorrected PSF halo in Stokes $I$
will be far larger than this statistical bias.  Recall our goal is not
to make absolute polarization measurements; when
attempting to use polarimetry to detect fainter signals hidden in the speckled 
PSF halo, it is simply infeasible
to make absolute measurements of $P/I$ and so we do not try.

Along with the mosaics, we generate exposure maps and
per-pixel empirical uncertainty images
for $I$, $Q$ and $U$, 
computed from the standard deviation of double-differenced pairs of
input frames.  
Computing the per-pixel uncertainty based on unsubtracted or
single-subtracted
images would not provide an accurate assessment of the
speckle-noise-reduced characteristics of the final Stokes images. 
During the
course of our AO polarimetry survey of Herbig Ae/Be stars, these uncertainty maps 
proved very 
valuable in assessing the
significance of faint signals.

\subsection{Final Steps}

The last steps in reduction are 
updating the World Coordinate System header, and 
flux calibration 
in Jy arcsec$^{-2}$ based on the
closest-in-time photometric standard. 
The final images are stored as multi-extension FITS files containing
the Stokes data cubes, exposure maps, and uncertainty images.

These data may be visualized using several IDL plotting routines, or
interactively explored using a version of the \texttt{atv} image
display tool \citep{2001ASPC..238..385B} modified to accept Stokes data cubes and overplot
polarization vectors. This code is available from the author's web
site\footnote{\texttt{http://astro.ucla.edu/$\sim$mperrin/idl/}}.

\section{Achieved Performance}\label{performance}

\subsection{Contrast Enhancement}

\begin{figure*}[t]
\begin{center}
\includegraphics[height=3.0in]{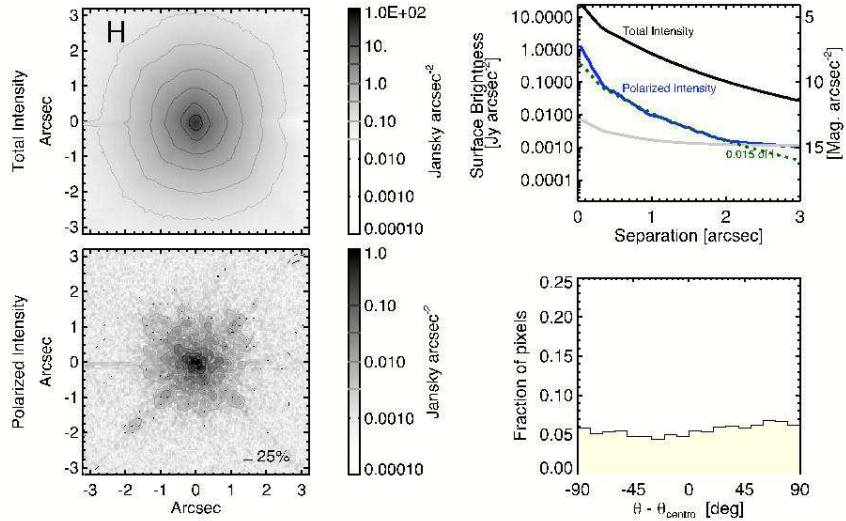}
\end{center}
\caption{
\label{image_HD18803}
$H$ band imaging polarimetry of the unpolarized standard HD 18803,
demonstrating the cancellation of unpolarized starlight.
\textit{Left:} Total and polarized
intensity. 
The display scale for polarized intensity is reduced by 100
compared with total intensity.
\textit{Right, top:} Radial profiles showing total and polarized intensity and the 
minimum theoretical noise floor
for $P$ based on photon statistics and read noise. 
As a rough estimate of the light-suppression capabilities
of the polarimeter, the thin dashed line shows a copy of the $I$
profile that has been multiplicatively scaled by $0.015$ to match the $P$
profile. In other words, the polarimeter rejects 98.5\% of unpolarized
starlight for this observation.
\textit{Right, bottom: } A histogram of polarization
position vectors relative to the expected
centrosymmetric pattern that indicates circumstellar dust (see \S
\ref{histogram-intro}). The
uniform
distribution seen here is the expected null result for an unpolarized star.  
}
 \end{figure*}

In practice the IRCAL polarimeter and data reduction pipeline remove 98-99\% of 
unpolarized starlight, greatly attenuating the background
against which scattered light must be detected (Figure \ref{image_HD18803}). The exact level of
performance varies with conditions.

\begin{figure}[htb]
\begin{center}
\includegraphics[height=2in]{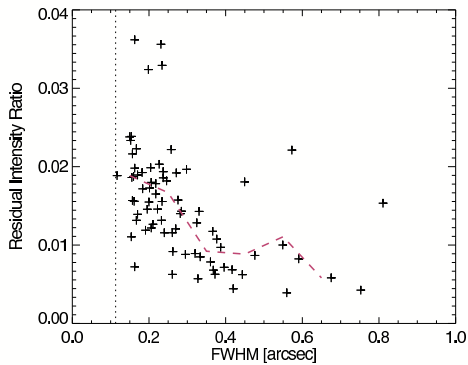}
\end{center}
\caption[Unpolarized Light Suppression Performance]{
\label{suppression}
The fraction of residual starlight not suppressed by polarimetry (measured as 
the ratio of apparent polarized intensity over
total starlight) plotted versus PSF full-width at half maximum (FWHM). The
dashed line shows the average suppression ratio in each 0.1\arcsec\ wide
bin, and the vertical dotted line shows the minimum diffraction-limited FWHM.
Higher ratios mean more unpolarized starlight is leaking through into
the final polarized intensity images.
The suppression of unpolarized starlight is best for
observations with large FWHM, and conversely worst when the AO
correction provides the best FWHMs. This surprising result indicates that high spatial
frequencies in the PSF contribute strongly to the subtraction
residuals, suggesting that registration and aliasing 
rather than flat fielding errors set the greatest limits on
performance for our polarimeter.

}
\end{figure}

Outside of $\sim$1\arcsec, we typically reach a noise floor $P_{noise}$ set by photon and
read noise. Inside that radius, instrumental systematics raise the
polarized surface brightness background above the theoretical noise floor. The
precise transition between photon-noise-limited and
systematic-limited depends on factors including source
brightness and seeing; for some of our targets the transition
occurs inside 1\arcsec\ (though rarely inside
0.7\arcsec) and for some it's further out.
This performance is similar to that achieved by
\citet{PotterPhD} with the Hokupa'a polarimeter.

Like speckle noise, the systematic noise floor is not Gaussian and does not improve with time as
$\sqrt{t}$. Additional integration increases the sensitivity at small
radii only
slightly if at all, similar to the behavior of semi-static speckles
in conventional AO imaging \citep[e.g.][]{Racine99,2004ApJ...612L..85A,Hinkley2007}. But we stress
here that the polarimetric noise floor is not merely semi-static speckles
rearing their ugly heads: a perfect polarimeter would completely
reject any static speckles, no matter how aberrated the $I$ PSF. It is
imperfections in the polarimeter that allow a fraction of those
speckles to bleed from $I$ into $P$ and ultimately limit our
sensitivity.

Many factors contribute to this imperfect subtraction of images. Is it possible to
disentangle which of these---flat fielding error, pupil shear,
registration error, and so on---is dominant? 
In practice, for IRCAL it has proven difficult to quantitatively assess the
relative importance of the various error terms. In part this is because of
the instrument's history. It was a mode
retrofitted into the camera at the telescope, and it never 
underwent any laboratory tests with high-quality polarized calibration sources.
More fundamentally, the various terms in the
error budget all
have similar functional forms, $P_{noise} \propto I$.
Flat field errors,
registration errors, pupil shear, all introduce noise
proportional to total intensity, rendering it hard to empirically disentangle the magnitude of
each individual effect. 

We thus present only a
partial resolution to this issue. For each target observed during 
our Herbig Ae/Be survey, we estimated the polarized noise
floor as a fraction of total intensity,
by computing the median ratio of
the radial profiles of $P$ and $I$ between 0.2\arcsec\ and
2.0.\arcsec\ 
In Figure \ref{suppression}, we plot this ratio against the PSF FWHM, a measure of wavefront quality.
The result is striking: as AO
performance improves (i.e. the FWHM becomes smaller), on average a larger fraction of
light ``leaks through'' the polarimeter.  
Better seeing and higher Strehl PSFs perversely result in slightly worse 
PSF subtractions!

From this we conclude that the subtracton residuals increase when the
PSF has more power at high spatial frequencies. Errors in flat
fielding would not cause this; such errors should be a simple
multiplicative factor on the PSF, regardless of PSF shape. On the
other hand, errors from PSF registration, aliasing, and numerical interpolation \textit{will} 
grow as high spatial frequency power increases, particularly
for an undersampled detector such as IRCAL's. 
Registration errors of saturated images will also contribute to this
problem.  Long
exposures are needed to detect faint outer nebulae, and in such data the core of
the PSF will be saturated and thus unusable for
registration. In conditions with better seeing, saturation will occur
earlier, contributing to the observed degradation. Thus for IRCAL it
appears that image registration and aliasing are a more significant
limitation than flat fielding.

\subsection{Detection of polarized light in reduced images}

In some cases polarized structures are immediately visible in reduced
images, but more often we face the task of detecting faint
polarizations at low signal to noise. How may we best decide whether a
given data set shows robust evidence for detected polarization or not?

\label{histogram-intro}

Merely looking by eye at polarization images is an
ineffective way to approach these data.
Polarization is fundamentally
a vector quantity, and examining only one of $Q$ or $U$ at a time 
does not take full advantage of that richness. 
Hunting for a faint signal in $P$ is also
not the best approach: $P$ incorporates both Stokes $Q$ and
$U$, but its positive definite nature biases it so any noise whatsoever 
results in a nonzero, positive value, which swamps faint polarized signals.
Numerical fits should be
done directly on the $Q$ and $U$ images, and only transformed to $P$
afterwards if necessary.

Luckily, light scattered from circumstellar dust has a distinct
signature we can search for. This characteristic may be
visualized either as the so-called ``butterfly'' pattern in the $Q$
and $U$ images, or as a ``centrosymmetric'' circular pattern of polarization vectors
arranged circumferentially around the central source.
\citet{2004AnA...415..671A} fit a sinusoid to the butterfly pattern in
$Q$ and $U$ images to 
detected scattered light to within
0.1\arcsec\ of TW Hya. 
A closely related technique uses instead the polarization position angle
$\theta$.  The position angles for centrosymmetric scattering are given by
\[
    \theta_{centro}(x,y) = \arctan \left( \frac{y-y_\ast}{x-x_\ast}
    \right),
\]
where $(x_\ast, y_\ast)$ is the central star's location. 
By comparing the
observed position angle $\theta_{obs}(x,y)$ with 
the expected position angle $\theta_{centro}$,
we can sensitively discriminate dust-scattered light from other contaminants
such as residual speckle noise or registration errors.  
A key advantage is that the position angle $\theta_{obs}$
is formed from the
ratio of $Q$ and $U$, and hence is less sensitive to flat fielding
errors than either of those quantities alone
\citep{PotterPhD}. 

\begin{figure*}[t]
\begin{center}
\includegraphics[height=3.0in]{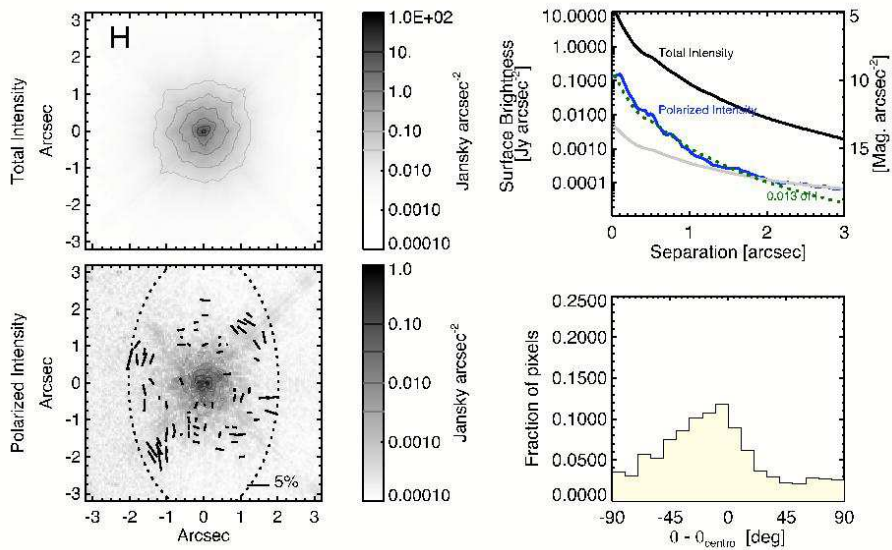}
\end{center}
\caption{
\label{image_HD141569} $H$ band polarimetry of a very faint signal: the circumstellar 
disk of the Herbig Ae star HD 141569. 
See the caption of Figure \ref{image_HD18803} for a description of the
various quantities plotted.  The dotted ellipse traces the outer
bright ring in the disk, as seen by ACS coronagraphy
\citep{2003AJ....126..385C}.
In these IRCAL data, little to no sign of circumstellar dust
scattering is apparent in the images or the radial profiles. However,
the histogram of position vectors has a definite peak near zero,
indicating that the majority of pixels above the polarized S/N
threshhold have centrosymmetric polarization angles, as expected for
dust scattering.
While these data do not provide much scientific
insight, this is a statistically significant detection of polarized
light, demonstrating the ability of differential polarimetry to
sensitively detect very faint signals for subsequent followup.
}
\end{figure*}

See Figure \ref{image_HD141569} for a
demonstration of using this method to detect a
very faint polarized signal from a disk around the Herbig Ae star HD
141569. Located at 99 pc, this young star has a circumstellar disk
believed to be transitioning from optically thick to optically thin,
possibly due to the clearing effects of unseen planets. Based on prior
coronagraphic observations
\citep{1999ApJ...525L..53W,2003AJ....126..385C,2003ApJ...585..494B}, the disk's $H$-band surface brightness
is 16-17 mag. arcsec$^{-2}$ between 1-2\arcsec radii, compared
to a stellar magnitude of $H=6.8$.  We observed HD 141569 on 2003 June
17 for 1600 s total, as part of a large survey of dust around Herbig Ae/Be stars
(Perrin et al\., in preparation).
The disk's structure is not at all apparent in the $I$ or $P$ images.
However, if we plot a histogram of the position angles for pixels\footnote{In addition to masking out pixels with low S/N, we also mask out
the diffraction spikes fixed at 45\degr when computing these
centrosymmetry histograms.}
which have polarized S/N $> 3$, we find they are predominantly close to
the expected centrosymmetric angles (seen in the lower right panel in Figure
\ref{image_HD141569}).
The observed polarization signal is strongest near the
location of the bright outer ring previously seen by NICMOS and ACS.
While this observation only marginally detects
this disk and does not reveal its structure clearly, 
to our knowledge this is the first non-coronagraphic
detection of HD 141569's disk in the near-IR.
This demonstrates that
centrosymmetry histograms can be used to sensitively detect extremely
faint polarized signals which might otherwise be missed.

Recently, \citet{2006A&A...452..657S} have advocated the use of ``radial Stokes parameters'' 
as an alternative technique that combines
aspects of both the butterfly and centrosymmetry methods. In this
approach the Stokes vectors are transformed according to 
\begin{eqnarray*}
    Q_r &=& +Q \cos 2\phi + U \sin 2\phi \\
    U_r &=& -Q \sin 2\phi + U \cos 2\phi 
\end{eqnarray*}
where $\phi$ is the position angle measured from the central source.
$Q_r$ is then a measure of the polarization's 
centrosymmetry.
The use of these quantities for polarimetric analysis deserves further study.

\begin{figure}[htbp]
\begin{center}
\includegraphics[height=3in]{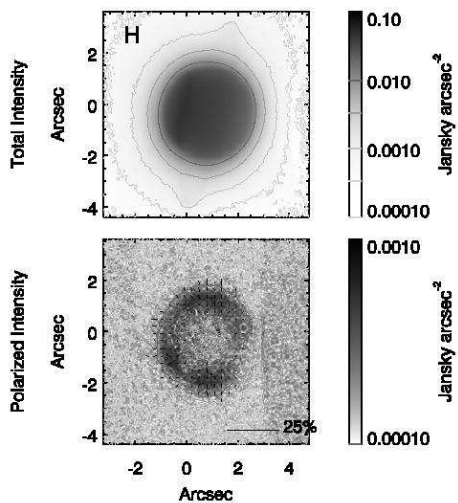}
\end{center}
\caption[AO Imaging polarimetry of Uranus]{
\label{image_Uranus}
Uranus, as seen in polarized $H$ band light on 2004 July 05. The upper
frame shows total intensity while the lower frame shows
polarized intensity with overplotted vectors giving the percent
polarization. Uranus is very limb-brightened in polarized light, with
a peak polarization of about 1.5\% at the edge of its disk. Unlike
circumstellar disks, the polarization vectors here point radially
inwards rather than circumferentially, an expected effect of
Rayleigh scattering in planetary atmospheres. The northern hemisphere, at right, is tilted 
away from us and
noticably less polarized than the nearer southern hemisphere. To our knowledge, this is
the first resolved imaging polarimetry of Uranus in the
near-infrared, and only the second imaging polarimetry of the planet at any
wavelength, after \citet{2006A&A...452..657S}.}
 \end{figure}

\begin{figure*}[htbp]
\begin{center}
\includegraphics[height=3in]{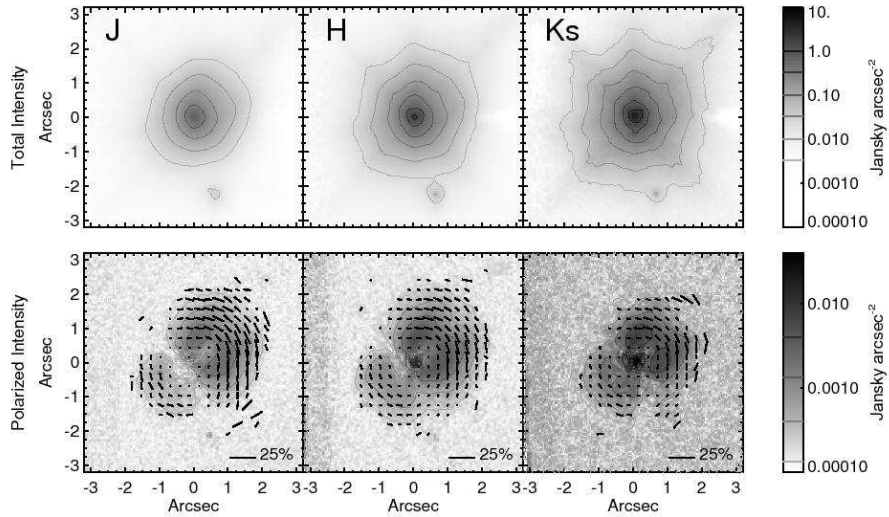}
\end{center}
\caption[\hspace{0.1in}AO Imaging polarimetry of Hen 3-1475]{
\label{image_PDS_465}
Multiwavelength imaging polarimetry of the pre-planetary nebula Hen
3-1475.  This data shows the power of polarimetry to reveal faint circumstellar material even in 
the presence of complicated and variable AO PSFs. In total
intensity (top row) the circumstellar nebula is not easily visible,
but in polarized intensity (bottom row) the symmetric bipolar nebula
is clearly detected in all three wavelengths. The morphology seen here
compares well with recent HST NICMOS observations presented by 
\citet{2007AJ....133.1345U}.
}
\end{figure*}

\begin{figure*}[htbp]
\begin{center}
\includegraphics[height=2.8in]{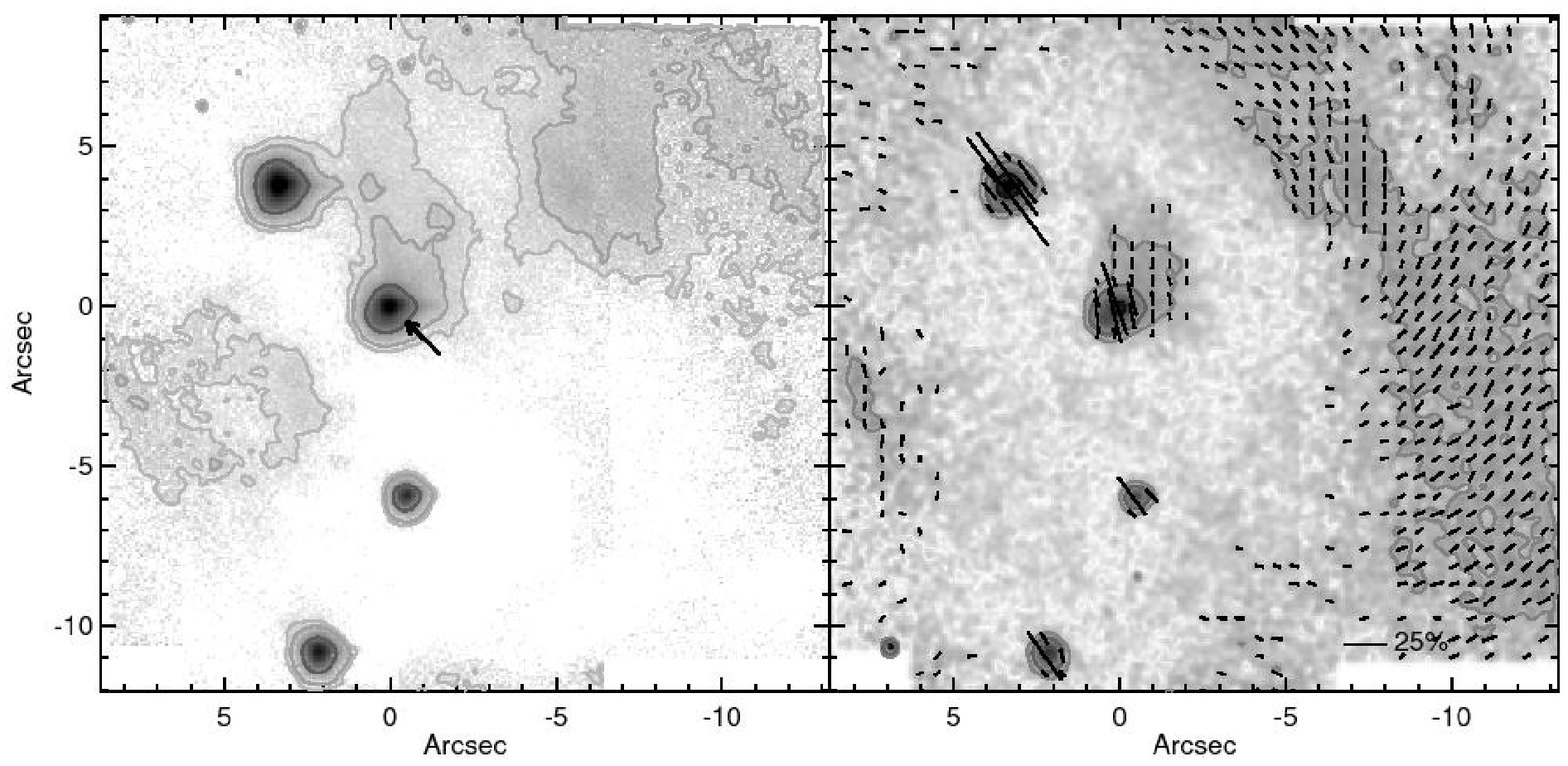}
\vskip -0.2in
\end{center}
\vskip -0.2in
\caption{
\label{image_crab}
Laser guide star AO imaging polarimetry of the inner portion of the Crab Nebula, M1, in $H$ band.  AO polarimetry can be
applied to observe sources of polarization besides dust scattering. 
Here we observe near-IR synchrotron emission from the Crab Nebula. 
The arrow indicates the Crab's central pulsar.
Several well known wisps and the knot immediately southeast of the
pulsar are visible. The observed morphology and polarization angles
are generally consistent with the higher angular resolution
observations obtained recently with HST ACS by
\citet{2007AAS...21110026H}. 
Measurement of the degree and orientation of polarization can provide 
insight into the magnetic field geometry and relatavistic outflow
properties. The large field of view shown here was obtained by
mosaicing together dithered images comprising 5700 s total
exposure time, taken on 2005 Nov 21.
}
\end{figure*}

\section{Conclusions} \label{conclusions}
   
   We have developed a dual-channel imaging polarimetry mode
   for the Lick Observatory adaptive optics system and its IRCAL
   science camera.    Its unique YLF Wollaston prism enables
   differential
   polarimetry throughout the near infrared, 
   enhancing contrast for the detection of faint circumstellar dust.
   In practice this system can reduce the stellar PSF
   halo by a factor of 50-100, 
   ultimately limited by instrumental systematics such as
   registration and flat-fielding errors. These are the same
   systematics which have limited other AO polarimeters such as
   at Gemini \citep{2003PhDT........20P} and the VLT
   \citep{2004AnA...415..671A}, and we report similar achieved performance to
   those instruments.

   After more than five years of operations, the IRCAL polarimeter has
   observed targets ranging from young stars to
   planets to planetary nebulae.  
A few observations serve to show the 
capabilities of the IRCAL polarimeter in a variety of scientific
contexts. 
Figure \ref{image_Uranus} presents the
first-ever $H$ band imaging polarimetry of Uranus.
Figure \ref{image_PDS_465} shows high-contrast, multiwavelength polarimetry of the
circumstellar material around a pre-planetary nebula. 
Lastly, Figure \ref{image_crab} presents polarization
measurements for the synchrotron emission from the inner region of the
Crab Nebula, demonstrating that the IRCAL polarimeter can also be used for
studies of polarizing processes besides dust scattering.

While the IRCAL polarimeter performs well, it suffers from certain
instrumental limitations that could have been avoided had
polarimetry been considered during the
initial design of the camera. All too many polarimeters are born this
way, as post-facto retrofits to instruments originally intended as
pure imagers \citep{2005ASPC..343....3H}.  This lesson is particularly
relevant to current design efforts for ``Extreme AO'' high
contrast imagers: polarimetric performance
should be considered in trade studies early in the design process of
both AO systems and science cameras. Present AO polarimeters (in
\textit{every} instance retrofitted into existing instruments) have ably
demonstrated their ability to detect faint scattered light in the
presence of a bright stellar PSF---the potential for discovery using a
truly optimized AO polarimeter is tremendous.
In IRCAL's case, one of the major difficulties is accurately
registering undersampled or core-saturated images.
Future polarimeters should avoid this by having a well-sampled detector and
using ``satellite
PSFs'', essentially intentional ghost images, to allow the precise
registration of images with saturated (or coronagraphically
occulted) PSF cores. 

An even more promising technique would be to split the polarizations only
\textit{after} pixellating the image plane, entirely
eliminating the subpixel registration problem and partially mitigating the
effects of uncertain flat fields.
Several recent designs for high contrast imaging spectrographs use focal plane
lenslet arrays to chop the PSF into pixels prior to
wavelength dispersion, to  
reduce non-common-path optical errors \citep[e.g.][]{2004ApJ...615L..61M,2005PASP..117..745M}. 
Experiments to validate these designs are currently underway both in
the laboratory \citep{2006SPIE.6269E..77L} and on the sky with OSIRIS at Keck \citep{2006SPIE.6269E..42L}.
Recent results are promising, but differential refraction (both
atmospheric and in AO system dichroics) poses complications. However,
for polarimetric applications, differential refraction is much less
of an issue, and the ability to modulate the polarization provides 
robustness against non-common-path errors. Lenslet-based differential
polarimeters deserve careful study as part of the high contrast
astronomer's future toolkit. As an added bonus, such an instrument
would be immune to lateral chromatism, eliminating the need for exotic
materials such as YLF.

We end with a brief list of lessons learned from IRCAL, applicable to
the design of future high contrast polarimeters. 
Starlight suppression of about 2 orders of magnitude is attainable
with IRCAL and all its flaws. A goal of 3 orders of magnitude
suppression seems not unrealistic for future instruments. 

\begin{enumerate}
      \item Low repeatability of the aperture and filter wheel mechanisms 
	    in IRCAL complicates data taking and reduction, but is not
	    insurmountable. Any reasonable future design will improve
	    on IRCAL in this regard.
	\item Repeatability of mechanisms is
	    particularly important as there are stringent requirements
	    on flat fielding, and IRCAL's mechanism nonrepeatability
	    prevents the development of high signal-to-noise flat
	    libraries. 
    \item The Wollaston prism should ideally be located after the cold pupil, to prevent
	  pupil shear between the two polarized beams. 
    \item    To minimize instrumental polarization, the
	  waveplate should be located as far upstream as possible. However, 
	  optimized scaled subtraction offers an
	  effective way to empirically remove instrumental (and
	  interstellar)
	  polarization for applications where absolute polarimetry is
	  not required.
    \item Registration errors are a key component of the error
	  budget, particularly for undersampled data or 
	long exposures saturated on the central star. 
	The use of artifical ``satellite PSFs'' as astrometric references
	  may help solve this problem.
	      \item  Lenslet arrays hold great promise for future high contrast
	   polarimeters, largely eliminating the image registration
	   problem.  Post
	  lenslet array, the requirements on lateral chromatism are
	  tremendously relaxed, potentially greatly easing
	  requirements on the beamsplitter.
    \item For detecting very faint polarized signals, centrosymmetry histograms provide a useful 
	  detection metric, particularly when
	  coupled with numerical uncertainty maps computed as part
	  of the data reduction process.
\end{enumerate}

\acknowledgements

    We would like to thank Oliver Meissner and the staff of
    Onyx Optics for their efforts in fabricating our Wollaston prism, and
    in particular for providing us with two extra prisms.   Thanks also
    go to Don Gavel and Elinor Gates for their work in integrating the
    polarimetry mode into the Lick AO system, to Brian
    Bauman for useful discussions on optics, and to the entire
    staff of Lick Observatory, without whom these observations would
    never have been possible.  Michael
    Fitzgerald provided invaluable technical assistance with IRCAL over the years, for which
    MDP will be forever grateful. 
    This work was supported by the National Science
    Foundation Science and Technology Center for Adaptive Optics,
    managed by the University of California at Santa Cruz under
    cooperative agreement No. AST 9876783. MDP was partially supported by a
    NASA Michelson Graduate Fellowship, under contract to the Jet
    Propulsion Laboratory
    (JPL), and is now supported by an NSF Astronomy and
    Astrophysics Postdoctoral Fellowship.

\appendix

\section{An updated list of materials for infrared Wollaston prisms}

\begin{figure}[htbp]
\begin{center}
\includegraphics[height=2.5in]{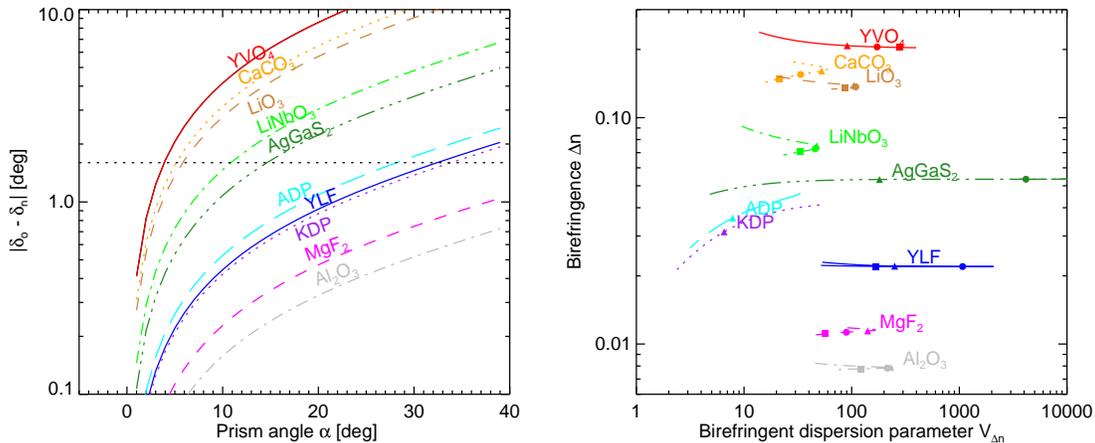}
\end{center}
\caption{ \label{material_plot}
\textbf{Left:} Beam separation angle as a function of Wollaston prism
angle, for birefringent materials from Table \ref{wollmats}. More
birefringent materials are higher in this plot, and require a smaller
prism angle to acheive a given beam separation.  The horizontal
line marks the desired separation for IRCAL. \textbf{Right:}
Birefringence versus birefringent dispersion for these materials. This
is the birefringent version of the standard Abbe plot for
optical materials.
On the horizontal axis, the birefringent dispersion parameter $V_{\Delta n}$  measures
how much birefringence varies with wavelength, with higher numbers
indicating less variation (see \S \ref{materials}). 
Imaging polarimeters benefit from materials whose birefringence is
more constant with wavelength (high $V_{\Delta n}$), such as YLF and
AgGaS$_2$. For each material, we show the change in these birefringent parameters
from 1-2.5 $\mu m$. Triangles, circles, and squares denote the central wavelengths of $J$, $H$, and $K_s$ respectively. 
}
\end{figure}

A large number of birefringent optical materials can be used to make
astronomical Wollaston prisms. \citet{Oliva97} provides a useful list
of some of the best candidates. Since that paper, updated measurements
of the optical properties of many of these materials have become
available, while other candidate materials have come to light. Table
\ref{wollmats} lists potential materials for near infrared Wollaston
prisms, and provides updated values for the birefringence $\Delta n$ and birefringent
dispersion parameter $V_{\Delta n}$, while Figure \ref{material_plot}
shows both achievable beam separations and birefringent optical
properties for there materials.

\begin{deluxetable}{llrrrrl}
\tablecolumns{7}
\tablecaption{Materials for Infrared Wollastons\label{wollmats}}
\tablewidth{0pt}
\tablehead{
Material & $n$  & $\Delta{n}$ & $V_{\Delta{n}}, J$ & $V_{\Delta{n}}, H$ &$V_{\Delta{n}}, K$ & References \\
}
\startdata
GaSe 		& 2.74 &  0.3195 &      157 &      135 &      484 & 1 \\
Ag$_3$As$_3$ 	& 2.77 &  0.2231 &       41 &       72 &      108 & 2 \\
YVO$_4$ 	& 2.15 &  0.2053 &       97 &      172 &      278 & 3  \\
CaCO$_3$ 	& 1.63 &  0.1550 &       50 &       33 &       21 & 4 \\
Ag$_3$SBs$_3$ 	& 2.83 &  0.1424 &       37 &       66 &      102 & 2 \\
LiO$_3$ 	& 1.85 &  0.1367 &      107 &      109 &       86 & 2 \\
BBO 		& 1.64 &  0.1158 &     1367 &      175 &       87 & 2\\
LiNbO$_3$ 	& 2.21 &  0.0728 &       48 &       45 &       33 & 2\\
AgGaS$_2$ 	& 2.42 &  0.0535 &      227 &     4152 &    11210 & 5, 6 \\
BABF 		& 1.61 &  0.0397 &       19 &       14 &        9 & 7  \\
ADP 		& 1.48 &  0.0261 &        7 &        - &        - & 8\\
YLF 		& 1.44 & -0.0220 &      268 &     1072 &      169 & 9  \\
KDP 		& 1.48 &  0.0208 &        5 &        - &        - & 10 \\
MgF$_2$ 	& 1.37 & -0.0113 &      135 &       89 &       56 & 11  \\
Al$_2$O$_3$ 	& 1.74 &  0.0078 &      235 &      213 &      122 & 12  \\
KTiOAsO$_4$ 	& 1.79 & -0.0055 &       56 &       26 &       17 & 1 \\
\enddata
\tablecomments{Materials are ordered by decreasing birefringence. The index of refraction, $n$, and birefringence, $\Delta{n}$, are stated for 1.65 \um. The polarization dispersion parameters $V_{\delta{n}}$
were calculated assuming filter bandpasses of 20\%, centered on 1.25,
1.65, and 2.1 \um. Dashes indicate wavelength ranges where accurate
index of refraction information was not available. More detailed
tables of optical properties and software implementing Sellmeier
models for all these materials is available from the authors.}
\tablerefs{
1: \citet{Allakhverdiev:2005p979}
2: \citet{Weber:2003p986}
3: \citet{Lomheim:1978p1021}
4: \citet{Ghosh:1999p739}
5: \citet{Willer:2001p386}
6: \citet{Takaoka:1999p378}
7: \citet{Hu:2002p1247}
8: CASIX data sheet, http://www.u-oplaz.com/crystals/crystals07.htm
9:  \citet{Barnes:1980p902}
10: Redoptronics data sheet, http://www.optical-components.com/KDP-crystal.html
11: \citet{Tropf:1995p381}
12: \citet{Kaplan:2003p377}
}
\end{deluxetable}

\clearpage
\bibliographystyle{apj-short}
\bibliography{apj-jour,pol,haebes,speckles,./optical_materials}

\end{document}